\documentclass{PoS}

\title{Black hole spin down in GRB observations and Cosmology}

\ShortTitle{Black Hole Spin Down}

\author{\speaker{Antonios Nathanail}\\
        Research Center
for Astronomy and Applied Mathematics,\\ Academy of Athens, Athens
11527, Greece \\ and \\
Section of Astrophysics, Astronomy and Mechanics,
Department of Physics,\\ University of Athens,
Panepistimiopolis Zografos, Athens 15783, Greece \\
        E-mail: \email{antonionitoni@hotmail.com}}
\author{Ioannis Contopoulos\\
        Research Center
for Astronomy and Applied Mathematics,\\ Academy of Athens, Athens
11527, Greece \\
        E-mail: \email{icontop@academyofathens.gr}}
\author{Spyros Basilakos\\
       Research Center
for Astronomy and Applied Mathematics,\\ Academy of Athens, Athens
11527, Greece \\
        E-mail: \email{svasil@academyofathens.gr}}

\abstract{According to Blandford \& Znajek (1977), energy can be extracted
electromagnetically from a rotating black hole, should the latter
be endowed with a magnetic field supported by  electric
currents  in a surrounding  disk. We show that exact models of black hole magnetospheres produce Poynting flux that
decreases almost exponentially with time. We went through the
Swift BAT-XRT lightcurves and identified a subclass of GRBs that
exhibits a clear exponential decay over more than three orders of
magnitude in flux (EDOHS GRBs). We estimate the energy  given-off
in the X-rays and discuss a possible correlation between the peak
brightness of the X-ray prompt emission and its decay time. We
  investigate a possible application of this result in high redshift
Cosmology.}

\FullConference{Swift: 10 Years of Discovery,\\
		2-5 December 2014\\
		La Sapienza University, Rome, Italy }

\begin{document}

\section{Introduction}

Gamma-ray bursts (hereafter GRBs) have been a great scientific
puzzle since their discovery in the 60s .
For more than 20 years, the only information we had about them was
that they produced flashes of gamma-rays. Their isotropic
distribution in the sky suggested a cosmological origin ,
and this was confirmed with the detection of the
afterglow in optical and other wavelengths that allowed the
determination of their redshift.
It has been widely discussed that GRBs fall into two
subcategories so-called short- and long-duration ,
which are believed to be associated with neutron
star-neutron star mergers and black hole formation during
super-massive star core collapse respectively.

 A
great amount of theoretical work has been invested in order to
understand what is the central engine and the emission mechanism
of GRBs. Several researchers believe that the central engine is a
black hole (for example \cite{W93}).
In recent years the question of the central engine has been put
aside, while research  focuses on the emission
region, the emission mechanisms and the effort to understand all
the characteristics of the light curves and the spectra of the
bursts (for a recent review \cite{KZ14}). The idea
of a black hole powering the burst is widely accepted for the
so-called long-duration GRBs, where the source of the burst is
believed to be the electromagnetic extraction of energy from the
black hole rotation (\cite{BZ77}). Obviously, as the black hole loses energy, it will spin
down. The question is at what rate.  \cite{CNP14}
 calculated in detail the black hole
electromagnetic spindown rate and showed that the extracted
Poynting flux decreases almost exponentially with time (see next
section).

GRBs may also be of fundamental importance for Cosmology
since they are observed up to very high redshifts at which the
distance modulus is most sensitive to the cosmological parameters.
This makes them ideal potential tracers of the Hubble relation if
we could somehow associate their absolute luminosity with one (or
more) of their observable parameters.
Several authors
have tested the cosmological implications of GRBs through several
empirical correlations between various properties of the prompt
and in some cases also the afterglow emission,
and their results have been eagerly
applied to constrain cosmological parameters .

In the present work we went through the Swift BAT/XRT data in
order to find a clear exponential decay of energy flux in time
since the trigger of the burst, which would indicate that we may
be following a black hole spinning down. We restrict ourselves
to GRBs with known redshift, so that we can
estimate the energetics of each burst. We identify a certain
subclass of 14 long duration GRB events which follow a
certain correlation between the peak brightness of the X-ray
prompt emission and its decay time. We discuss how this
correlation may be used in high redshift cosmology.

\section{Black Hole Electromagnetic Spin Down}

The extraction of the rorational energy of a black hole through
the Blandford \& Znajek (1977) mechanism is widely accepted and
discussed in the literature. As we said, extraction of
rotational energy from the black hole actually corresponds to
black hole spin down. We review here the main elements of black
hole spin down as obtained in detail in references  \cite{CNP14}.
Let us consider a
supermassive star whose core collapses and forms a maximally
rotating black hole. If the star is magnetized, magnetic flux will
be advected with the collapse. The material that is going to
collapse into a black hole will be strongly magnetized, and
therefore its core will pass through a spinning magnetized neutron
star stage. A certain amount of magnetic flux $\Psi_m$ is then
going to cross the horizon. An equatorial  thick disk will form
around the black hole due to the rotational collapse. This
material will hold the magnetic flux advected initially toward the
horizon and (at least for a limited time) will prevent it from
escaping to infinity. As long as this is the case, the black hole
will lose rotational/reducible energy at a rate
\begin{equation}
\dot{E} \approx -\frac{1}{6\pi^2 c}\Psi_m^2\Omega^2\ ,
\label{EdotIa}
\end{equation}
and will thus spin down very dramatically (see references in \cite{BZ77})
for low spin parameters; references in \cite{CKP13},
\cite{NC14} for maximally rotating black holes).
Moreover, some (yet unknown) fraction $f_{\rm X}$ of it will be
observed in X-rays, namely $\dot{E}_{\rm X}\sim f_{\rm X} \dot{E}\ $.


A note on the production of high energy radiation in the black
hole magnetosphere is in order here. We have found that a generic
feature of black hole magnetospheres is a poloidal electric
current sheet that originates on the horizon at the equator.
Our observational and theoretical experience from
pulsars suggests that high energy radiation is expected to
originate from reconnection processes that result in particle
acceleration along the magnetospheric current sheet (\cite{KHKC12}, \cite{SS14}).
This implies that high energy radiation may not be coming
along the axis of rotation but in a direction orthogonal to it.
Obviously, further out, the current sheet and the
consequent high-energy radiation will be naturally collimated
along the axis of rotation by the stellar material surrounding the
black hole.

Let us now return to the calculation of the black hole spin down.
The available rotational/ reducible black hole energy is $a {\cal
G}M^2 \Omega/c$ where $a\equiv J/M$ is the black hole spin
parameter , and ${\cal G}$ is the
gravitational constant. The black hole will therefore spin down as
\begin{equation}
\dot{E} = \frac{{\cal G}M^2}{c}\frac{{\rm d}(a\Omega/M)}{{\rm
d}t}\ ,
\label{EdotIb}
\end{equation}

Equating eqs. ~(\ref{EdotIa}) and (\ref{EdotIb}) we find an
approximate solution of the rate of the black hole spin down (the
exact solution can be found in \cite{CNP14})
\begin{equation}
\dot{E}\approx  \dot{E_o}e^{-t/\tau},
\label{Eapprox}
\end{equation}
where
\begin{equation}
\tau\equiv \frac{24 c^5}{{\cal G}^2 B^2 M}=7
\left(\frac{B}{10^{16}\ \mbox{G}}\right)^{-2}
\left(\frac{M}{M_\odot}\right)^{-1}\ \ \mbox{sec}\
\end{equation}
is a very important physical parameter of the model. Here, $B$ is the initial black hole magnetic field.
 As we will
discuss below, $\tau$ may be directly observable.


One can directly check that, for a magnetic field $ > 10^{15}
G$, the black hole spins down in a few tens to a few hundred
seconds. Obviously, such ultra-strong magnetic fields can survive
only during the core-collapse of a massive star, and are
subsequently dispersed away from the black hole horizon. We are
thus willing to make a {\em tentative association between GRB
events where eq.~(\ref{Eapprox}) yields a good fit in the decay of
the prompt emission
and the electromagnetic spin down of a newly formed
maximally rotating stellar mass black hole}.

\section{GRBs with Exponential Decay in One Hundred Seconds}

GRB prompt (and afterglow) emission light curves are in general
rather complicated with multi peak sub-structure. We want to find
signs of black hole spindown in GRB observations, that is why we
decided to focus only on GRBs with {\em a single exponentially
decaying prompt emission event}.
Furthermore, we  wanted to find similar systems so as to
explore and compare the energetics of the bursts.
We thus formulated the
following {\em empirical} criteria that characterize a particular
subclass of gamma-ray bursts:
1)  A single prompt emission event in 15 to 50~keV X-rays;
2) Exponential prompt emission decay over more than three
orders of magnitude; 3)  Prompt emission duration (up to the
first break where the energy flux seems to be constant)
longer than about 100 seconds, but not much longer
than a few hundred seconds ({\em ad hoc } value);
4)  Full (not sparse) sampling of the light curve during that
time interval;
5) Known redshift. This is required in order for us to test our
model of standard explosions for Cosmology.

\begin{table}
\begin{tabular}{lcccccr}
{\em Name} &{ \em z} &
$F_{\rm obs}(\pm \sigma_{F_{\rm obs}})$&
 $\tau_{\rm obs}(\pm \sigma_{\tau_{\rm obs}})$  &
  $E_X(\pm \sigma_{E_X})$\\
   &    & $10^{-8}\
\frac{\mbox{erg}}{\mbox{s}\ \mbox{cm}^2}$ & s
& $10^{51} \mbox{erg}$  \\ \\
130831A  &  0.4791&  $21(\pm 10)$     & $8.5(\pm  1)$    & $1.26 (\pm 0.45)$    \\
100418A   & 0.6235&  $8(\pm 1)$       & $6(\pm  1)$      & $0.33 (\pm 0.07)$    \\
080916   & 0.689  &  $7(\pm  1.2)$    & $10.1(\pm  2)$     & $0.83 (\pm 0.22)$    \\
061110A   & 0.757 &  $4.6(\pm 2)$     & $10.1(\pm  1.3)$   & $1.06 (\pm 0.48)$    \\
080430    & 0.767 &  $10.6(\pm 2)$    & $5.2(\pm  2)$    & $2.3 (\pm 0.98)$    \\
070714B   & 0.92  &  $5.4(\pm 2)$     & $29(\pm  2)$     & $3.68 (\pm 1.39)$    \\
051006   & 1.059  &  $6.56(\pm 1.4)$  & $7.5(\pm 0.7) $  & $1.31  (\pm 0.3)$    \\
110808A  &  1.348 &  $2.36(\pm 1)$    & $11(\pm  0.7)$   & $1.13 (\pm 0.48)$    \\
120724A  &  1.48  &  $2.1(\pm 0.7)$ & $17(\pm  1.4)$   &  \\
060708   & 1.92   &  $4(\pm 0.8)$     & $7(\pm  0.8)$    &   \\
070110   &  2.352 &  $2(\pm 0.9)$     & $21.4(\pm  2.1)$    &  \\
111107A  &  2.893 &  $0.5(\pm 0.3)$    & $27.5(\pm  3.5)$    &  \\
091109A  &  3.076 &  $2.4(\pm 1.2)$    & $23.5(\pm  3.1)$    &  \\
090519   &  3.85  &  $1.7 (\pm 0.5)$    & $39.4(\pm  4.5)$    &

\end{tabular}
\caption{EDOHS GRB Observations, 8 GRBs with $z < 1.4$ in the range of available SnIa
data , 2 GRBs with intermediate redshifts $1.4<z\leq 2$  and 4 GRBs with
high redshift $z>2$.}
\end{table}

The new GRB sub-class defined by the above criteria (hereafter
EDOHS GRBs) has similarities with previously defined sub-classes
such as Fast Rise Exponential Decay-FREDs selected by the Compton
GRO collaboration (\cite{KRL03}). We  worked with the Swift data and went through
all the Swift-BAT/XRT light curves from 2005 till 2014
(\cite{E07}, \cite{E09}). We found 14 GRBs that fulfil our
criteria and their BAT+XRT light curves can be fit with a simple
exponential (note that such a fit can be discerned more
clearly in a log-linear plot; see Appendix). For each one of
these events, the fit gave us the observed characteristic
exponential decay timescale $\tau_{\rm obs}$ and the peak observed
flux $F_{\rm obs}$ (Table~1).

The signs of exponential decay in our sample allow us to believe
that we may actually be observing the electromagnetic spin down of
the newly formed black hole. Furthermore, the deduced decay times
show a clear statistical time dilation with redshift (with only a
couple notable exceptions; see Fig.~1), which suggests
that they may indeed consist an unbiased representation of their
intrinsic values. Notice that this is not the case for the larger
GRB population where several observational selection biases
obscure the cosmological time dilation.
In other words there is some indication that our
selected GRB subclass (but not the entire GRB population) may
consist alternative cosmological distance indicators.

In the case of exponential decay that we are currently
investigating, the  total energy $E_{\rm X}$ given-off in X-rays
can be estimated as
\begin{equation}
E_{\rm X}=  \frac{4 \pi}{1+z} d_{\rm L}^2(z)F_{\rm obs}\tau_{\rm
obs}\ ,\label{EtotalX1}
\end{equation}
where, $d_{\rm L}$ is the luminosity distance.
We acknowledge that our
estimation is rather crude, and plan to study the spectra in a
future work.

\begin{figure}[t]

\includegraphics[trim=0cm 0cm 0cm 0cm,
clip=true, width=6.5cm, angle=0]{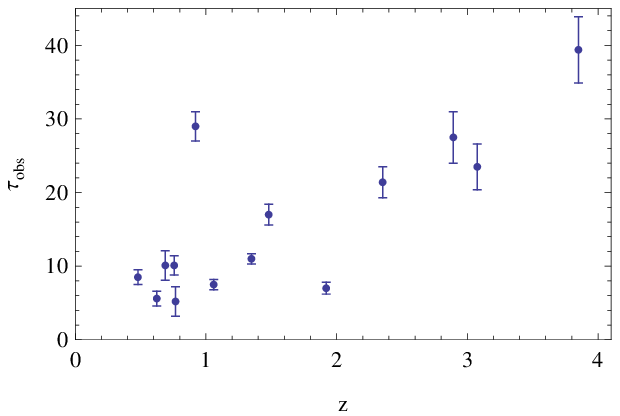}
\includegraphics[trim=0cm 0cm 0cm 0cm,
clip=true, width=8cm, angle=0]{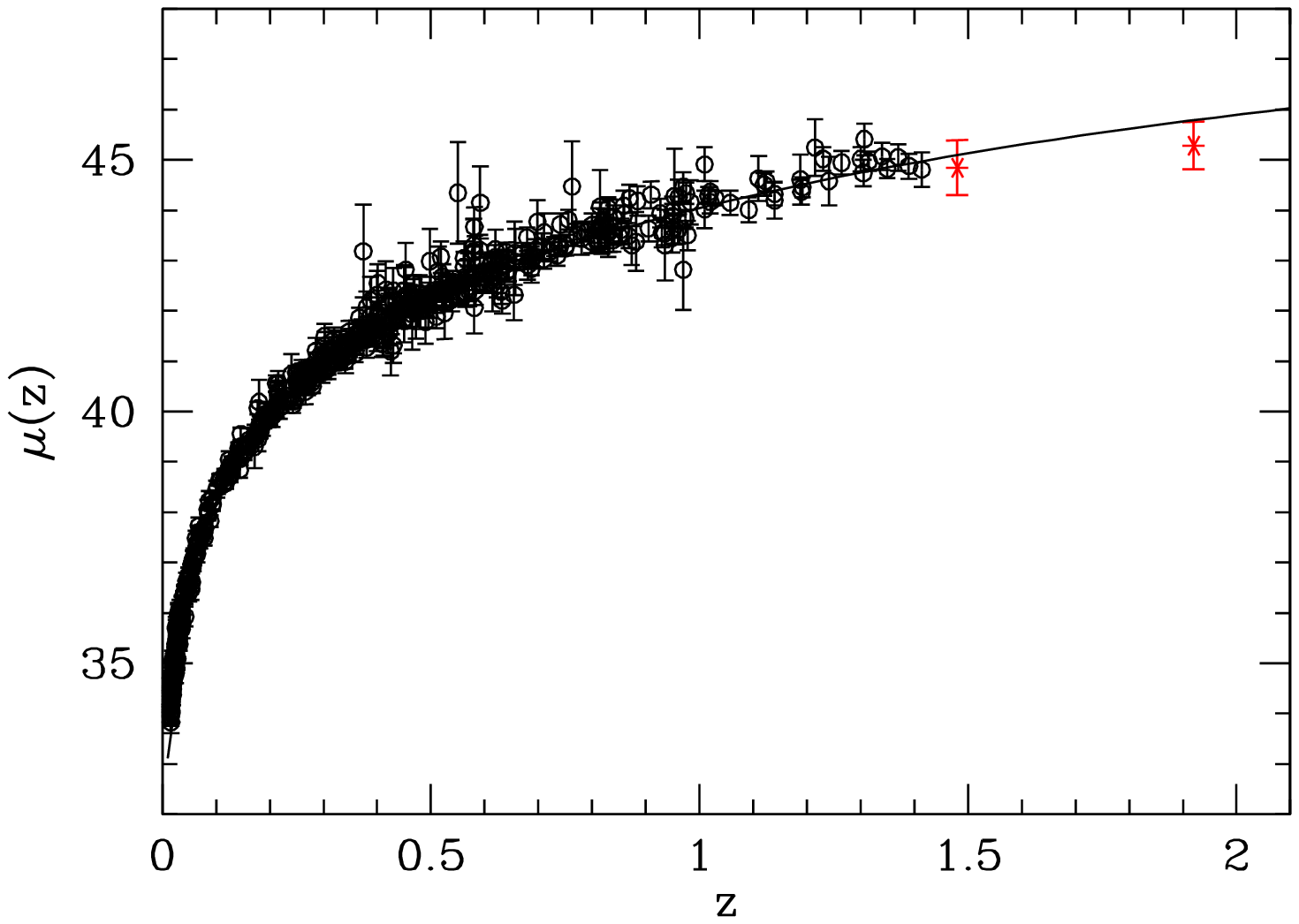}
\caption{Left:The observational characteristic exponential decay  timescale
$\tau_{obs}$ versus redshift $z$, signs of time dilation are evident in this plot.
Right:Comparison between the estimated distance modulus for the
intermediate EDOHS GRB observations ($1.4<z\le 2$, red stars)
and the recent SnIa data-set of \cite{S12} ($0.015\leq
z\leq 1.414$,  circles). The solid curve corresponds to the
concordance $\Lambda$CDM model with $\Omega_{m0}=0.30$ and
$H_{0}=70{\rm /km/s/Mpc}$.}
\end{figure}
We can estimate the luminosity distance of
EDOHS GRBs in the SnIa range ($0 < z < 1.4$) and calculate the
energy $E_{\rm X}$ given-off in the X-rays.
We tested whether
$E_{\rm X}$ depends on redshift. We find no significant
correlation between $E_{\rm X}$ and redshift in that range (a
$\chi^2$-minimization in $\log E_{\rm X}-\log(1+z)$ space yields an unacceptably high
reduced chi-square fit, $\chi_{\nu}^2 \gg 6.8$ and the probability
to fit the data is $\sim 10^{-7}$). The total X-ray energy release is
$
 E_{\rm X} \simeq 1.49(\pm 0.55)\times 10^{51}\
\mbox{erg}\ .
$

\section{Discussion and conclusions}

In our present study, we analyzed Swift-XRT/BAT light curves
(2005-2014) to find GRBs that fulfill a set of phenomenological
criteria defined in \S~3. In particular, we did a systematic
search for clear signs of exponential decay in the prompt and afterglow
emission, and obtained a new subclass of EDOHS GRBs.
We tentatively  associated  EDOHS GRBs with the electromagnetic spin
down of the black hole that forms during the core collapse of a
massive star. If that is indeed the case, then we may be probing
the Blandford-Znajek process in action.

We propose
to do high redshift cosmology with EDOHS GRBs.
Unfortunately, beyond redshift
$z\approx 2$ we are reaching the Swift-BAT trigger limit for EDOHS,
thus our current  high redshift events are not reliable cosmological tracers.
Future detectors will solve this problem.

This work made
use of data supplied by the UK Swift Science Data Centre at the
University of Leicester, and was supported by the General
Secretariat for Research and Technology of Greece and the European
Social Fund in the framework of Action `Excellence'.

\begin{figure}[t]
\includegraphics[trim=0cm 0cm 0cm 0cm,
clip=true, width=8cm, angle=0]{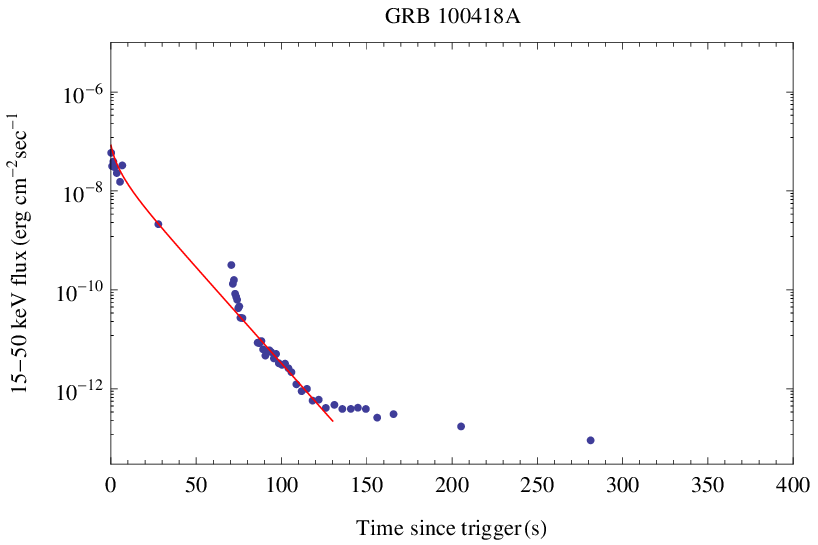}
\includegraphics[trim=0cm 0cm 0cm 0cm,
clip=true, width=8cm, angle=0]{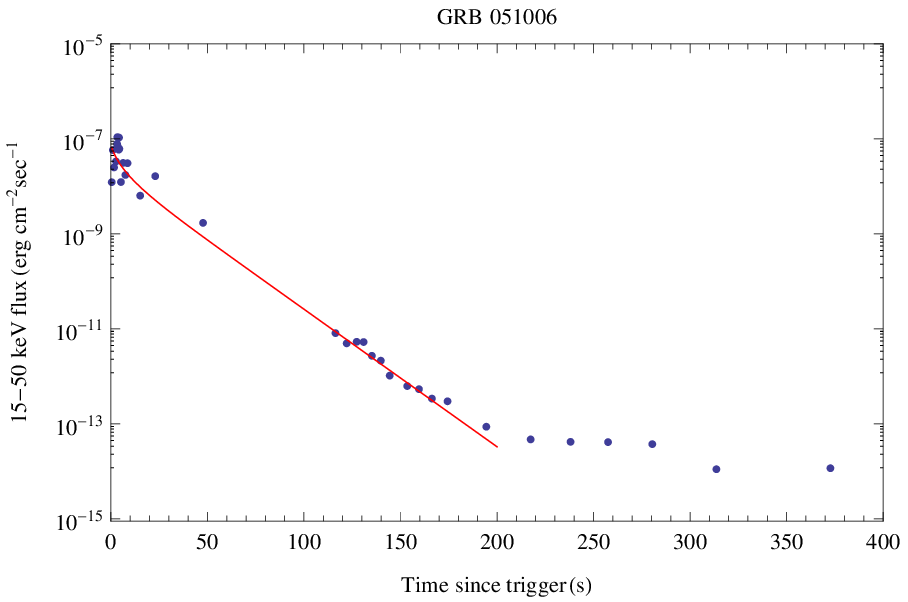}\\
\caption{Two of the selected GRBs with Exponential Decay in One
Hundred Seconds - EDOHS with  best fits of their
prompt emission almost exponential decay. The fits
yield the values of $\tau_{\rm obs}$ and $F_{\rm obs}$ shown in
Table~1.}
\end{figure}

\end{document}